\documentclass[letterpaper, 10 pt, conference]{ieeeconf}  
\newtheorem{remark}{Remark}

\IEEEoverridecommandlockouts                              

\usepackage{graphics,subcaption} 
\usepackage{epsfig,epstopdf} 
\usepackage{mathptmx} 
\usepackage{times} 
\usepackage{amsmath} 
\usepackage{amssymb}  
\usepackage{url}
\usepackage{color}
\title{\LARGE \bf
Multi-Agent Coverage Control with Energy
Depletion and Repletion*
}

\author{Xiangyu Meng$^{1}$, Arian Houshmand$^{1}$ and Christos G. Cassandras$^{1}$
\thanks{*This work was supported in part by NSF under grants ECCS-1509084,
IIP-1430145, and CNS-1645681, by AFOSR under grant FA9550-12-1-0113, by DOE
under grant DOE-46100, and by Bosch and MathWorks.}
\thanks{$^{1}$The authors are with Division of Systems Engineering, Boston University,
Brookline, MA 02446 USA {\tt\small xymeng@bu.edu; arianh@bu.edu; cgc@bu.edu}}%
}

\begin{document}

\maketitle
\thispagestyle{empty}
\pagestyle{empty}

\begin{abstract}
We develop a hybrid system model to describe the behavior of multiple agents cooperatively solving an optimal coverage problem under energy depletion and repletion constraints. The model captures the controlled switching of agents between coverage (when energy is depleted) and battery charging (when energy is replenished) modes.
It guarantees the feasibility of the coverage problem
by defining a guard function on each agent's battery level to prevent it from dying on its way to a charging station.
The charging station plays the role of a centralized scheduler to solve the contention problem of agents competing for the only charging resource in the mission space.
The optimal coverage problem is transformed into a parametric
optimization problem to determine an optimal recharging policy. This problem is solved through the use of Infinitesimal Perturbation Analysis (IPA), with simulation results showing that a full recharging policy is optimal.
\end{abstract}

\section{INTRODUCTION}
Systems consisting of cooperating mobile agents are often used to perform
tasks such as coverage \cite{zhong2011distributed, leonard2013nonuniform, kantaros2015distributed},
surveillance \cite{tang2005motion}, monitoring and sweeping
\cite{smith2012persistent}. A coverage task is one where agents are deployed
so as to cooperatively maximize the coverage of a given mission space
\cite{meguerdichian2001coverage}, where \textquotedblleft
coverage\textquotedblright\ is measured in a variety of ways, e.g., through
the joint detection probability of random events cooperatively detected by the
agents. Widely used methods to solve the coverage problem include distributed
gradient-based algorithms \cite{zhong2011distributed} and
Voronoi-partition-based algorithms \cite{cortes2004coverage}. These
approaches typically result in locally optimal solutions, hence possibly poor
performance. To escape such local optima, a boosting function approach is
proposed in \cite{sun2014escaping} where the performance is ensured to be
improved. Recently, the coverage problem was also approached by exploring the
submodularity property \cite{zhang2016string} of the objective function, and
a greedy algorithm is used to guarantee a provable bound relative to the
optimal performance \cite{sun2017submodularity}.

In most existing frameworks, agents are assumed to have unlimited on-board
energy to perform the coverage task. However, in practice, battery-powered
agents can only work for a limited time in the field \cite{leahy2016persistent}. For example, most
commercial drones powered by a single battery can fly for only about 15
minutes. Therefore, in this paper we take into account such energy constraints
and add another dimension to the traditional coverage problem. The basic setup
is similar to that in \cite{zhong2011distributed}. Agents interact with the
mission space through their sensing capabilities which are normally dependent
upon their physical distance from an event location. Outside its sensing
range, an agent has no ability to detect events. Unlike other multi-agent
energy-aware algorithms whose purpose is to reduce energy cost \cite{aksaray2016dynamic}, we assume that
a charging station is available for agents to visit according to some policy.
The objective is to maximize an overall environment coverage measure by
controlling the movement of all agents in a cooperative manner while
guaranteeing that no agent runs out of energy while in the mission space.

We provide a solution to the above problem by modeling the behavior of an
agent through three different modes: coverage (Mode 1), to-charging (Mode 2),
and in-charging (Mode 3). We assume that an agent has no prior knowledge of
the mission space except for the location of the charging station and the
positions of agents within its communication range. While in Mode 1, each
agent moves along the gradient direction of the objective function at the
maximum velocity so as to cooperatively maximize the coverage measure. As an
agent's energy\ is depleted, the agent switches to Mode 2 according to a guard
function designed to guarantee that a minimum energy amount is preserved to
reach the charging station from its current location while traveling at
maximum speed. Note that an agent shares its position and battery state
information with the charging station only when it is in the to-charging mode
(Mode 2). Since the charging station is shared by all agents, there can only
be at most a single agent at the station at any time. Therefore, two
scheduling algorithms are proposed to resolve contention among low-energy
agents: $(i)$ First-Request-First-Serve (FRFS), and $(ii)$
Shortest-Distance-First (SDF). These two scheduling algorithms are described
in detail in Section~\ref{MR}.
The charging station is perceived as a centralized controller executing a
scheduling algorithm by dictating agents' speeds so that a queue is formed by
agents while in Mode 2. In Mode 3, an agent is located at the charging station
and a model is developed for the battery charging dynamics using the dwell
time of an agent at the station as a controllable parameter to be optimized. The details for the modeling can be found in~\cite{meng2018hybrid}, and here we focus on using IPA in order to obtain the optimal dwell time of all agents at the charging station.

The contributions of this paper are summarized as follows. First, a hybrid
system model is developed so that the optimal coverage problem can be
transformed into a parametric optimization problem which can be subsequently
solved using IPA \cite{cassandras2010perturbation}. Second, two scheduling policies, FRFS and SDF, are proposed to allow agents
to share the charging station effectively while also guaranteeing that no
agent runs out of energy during the entire process.
Finally, we show that it is optimal to fully charge the battery when agents are in the in-charging mode.


The remainder of the paper is organized as follows. 
The optimal coverage
problem with energy depletion and repletion is formulated in Section~\ref{PF}
including the sensing model, charging and discharging dynamics of an agent. A
hybrid system model for the optimal coverage problem with energy depletion and
repletion is presented in Section~\ref{hm}, where we define guard functions to
control the switchings of an agent among different modes. Two scheduling algorithms are presented in Section~\ref{MR}. The solution of the optimization problem based on the
constructed hybrid system model is addressed in Section~\ref{IPAC}, followed by simulation examples in Section~\ref{SIM}. Concluding
remarks are given in Section~\ref{con}.

\section{PROBLEM FORMULATION}

\label{PF} Consider a bounded mission space $\Omega\in\mathbb{R}^{2}$, which
is modeled as a non-self-intersecting polygon. We deploy $N$ agents in the
mission space to detect possible events that may occur in it. By viewing the
position of agent $i$ in $\mathbb{R}^{2},$ its coordinates $s_{i}=\left[
x_{i},y_{i}\right]  ^{T}$ obey the following dynamics:%
\begin{align}
\dot{x}_{i}\left(  t\right)   &  =v_{i}\left(  t\right)  \cos w_{i}\left(
t\right)  ,\label{sd}\\
\dot{y}_{i}\left(  t\right)   &  =v_{i}\left(  t\right)  \sin w_{i}\left(
t\right)  , \label{sd1}%
\end{align}
with $v_{i}\left(  t\right)  $ denoting the speed, and $w_{i}\left(  t\right)
$ the heading direction of agent $i$. We assume that $v_{i}\left(  t\right)
\in\left[  0,v\right]  ,$ and $w_{i}\left(  t\right)  \in\left[
0,2\pi\right)  ,$ where $v$ is the maximum speed of an agent. The mission
space does not contain obstacles. If it does, the problem can be modified
appropriately as done in \cite{zhong2011distributed}.

In contrast to traditional multi-agent coverage problems, agents are assumed
to have a limited on-board energy supply, which is modeled by the
state-of-charge $q_{i}(t)$ of its battery (i.e., the fraction of the battery
available at time $t$). The dissipation of energy is proportional to a
quadratic function of the velocity, yielding the following dynamics:%
\begin{equation}
\dot{q}_{i}\left(  t\right)  =-\alpha v_{i}^{2}\left(  t\right)  , \label{bd}%
\end{equation}
where $\alpha$ is a scaling constant to ensure that $0\leq q_{i}(t)\leq1$.
When $q_{i}\left(  t\right)  $ is negative, this implies that agent $i$ is
\textquotedblleft dead\textquotedblright\ in the mission space.

\begin{remark}
The energy depletion model (\ref{bd}) is a simplified
version of the one used in \cite{setter2016energy} where the
agent motion is modeled by a double integrator and the energy dynamics are
modeled as%
\begin{equation*}
\dot{q}_{i}\left( t\right) =-v_{i}^{2}\left( t\right) -au_{i}^{2}\left(
t\right) ,
\end{equation*}%
where $v_{i}\left( t\right) $ is the velocity and $u_{i}\left( t\right) $ is
the acceleration. We assume that an agent's speed can be controlled
directly, therefore, we do not include the acceleration in (\ref{bd}). Here,
we also neglect energy costs associated with sensing and computation. The communication cost which depends on the distances from neighbor agents will be considered in future work based on:%
\begin{equation}
\dot{q}_{i}\left( t\right) =-\alpha v_{i}^{2}\left( t\right) -%
\eta \sum\nolimits_{j\in \mathcal{N}_{i}\left( t\right) }\left\Vert
s_{i}\left( t\right) -s_{j}\left( t\right) \right\Vert ^{2}\text{,}
\label{bd1}
\end{equation}%
where $\alpha$ and $\eta$ are two scalars, $\mathcal{N}_{i}$ is the set of neighbors of agent $i$ defined as%
\[
\mathcal{N}_{i}=\left\{  j\left\vert \Omega_{j}\cap\Omega_{i}\neq
\emptyset\right.  \right\}  ,
\]%
and $\Omega_{i}$ is the sensing range of agent $i$ to be defined later.
\end{remark}

To prevent agents from dying in the mission space, a charging station is
available to all agents to replenish their energy supply during the mission
time. Without loss of generality, we assume that the charging station is
located at the origin with coordinates $\left(  0,0\right)  $. At the charging
station, the charging process has the following dynamics:%
\begin{equation}
\dot{q}_{i}\left(  t\right)  =\beta, \label{b_charge}%
\end{equation}
where $\beta>0$ is the charging rate. We assume that only one agent can be
served at the charging station at any time.

Our objective is to maximize the coverage of the mission space $\Omega
\in\mathbb{R}^{2}$ over a time interval $[0,T]$ with $T$ being the time horizon, and at the same time keep all
agents alive, that is, $q_{i}\left(  t\right)  >0$ for all $t\in\lbrack0,T]$.
The case $q_{i}\left(  t\right)  =0$ can occur only at the charging station
$\left(  0,0\right)  $. Therefore, we consider the following optimization
problem for each agent $i$:%
\begin{equation}%
\begin{tabular}
[c]{cl}%
$\underset{w_{i}\left(  t\right)  ,\text{ }v_{i}\left(  t\right)  }{\max}$ &
$\frac{1}{T}\int_{0}^{T}H\left(  \mathbf{s}\left(  t\right)  \right)  dt$\\
s.t. & $q_{i}\left(  t\right)  \geq0,$\\
& $q_{i}\left(  t\right)  >0$ when $s_{i}\left(  t\right)  \neq\mathbf{0}$,\\
& (\ref{sd}), (\ref{sd1})\\
& $0\leq v_{i}(t)\leq v$,\\
& (\ref{b_charge}) if charging, (\ref{bd}) otherwise\\
& if $s_{i}(t)=\mathbf{0}$,\\
& then $s_{j}(t)\neq\mathbf{0}$ for all $j\neq i$\\
& $i=1,\ldots,N,$%
\end{tabular}
\ \label{of}%
\end{equation}
where $\mathbf{s}=[s_{1}^{T},\ldots,s_{N}^{T}]^{T}$ is a column vector that
contains all agent positions, and $H\left(
\mathbf{s}\left(  t\right)  \right)  $ is the coverage metric. We adopt the
coverage objective function used in \cite{zhong2011distributed} by first
defining a reward function $R\left(  x,y\right)  $ with $\left(  x,y\right)
\in\Omega$ to capture the \textquotedblleft value\textquotedblright\ of a
point $\left(  x,y\right)  $ in the mission space, and assume%
\[
\int\int_{\Omega}R\left(  x,y\right)  dxdy<\infty\text{.}%
\]
Thus,\ $R\left(  x,y\right)  $ may have larger values for points whose
coverage may carry more significance. Clearly, if all points in $\Omega$ are
treated indistinguishably, then $R\left(  x,y\right)  =1$ for all $\left(
x,y\right)  \in\Omega$.

Each agent has an isotropic sensing system with range $\delta_{i}$, that is,
an agent is able to cover the area
\[
\Omega_{i}\left(  x_{i},y_{i}\right)  =\left\{  \left(  x,y\right)  |\left(
x-x_{i}\right)  ^{2}+\left(  y-y_{i}\right)  ^{2}\leq\delta_{i}^{2}\right\}
\text{.}%
\]
The sensing probability of an agent at a point $\left(  x,y\right)  $ within
its sensing range $\Omega_{i}\left(  x_{i},y_{i}\right)  $ is characterized by
the sensing function $p_{i}\left(  x,y,x_{i},y_{i}\right)  \in\left[
0,1\right]  $ and depends on the distance between the agent location $\left(
x_{i},y_{i}\right)  $ and the point $\left(  x,y\right)  $. In particular, it
is monotonically decreasing in the distance between $\left(  x_{i}%
,y_{i}\right)  $ and $\left(  x,y\right)  $ and if a point $\left(
x,y\right)  $ is out of the sensing range of agent $i$, that is, $\left(
x,y\right)  \notin\Omega_{i}\left(  x_{i},y_{i}\right)  $, then $p_{i}\left(
x,y,x_{i},y_{i}\right)  =0$. For any given point $\left(  x,y\right)  $ in the
sensing range of multiple agents, assuming independence among agent sensing
capabilities, the joint event detection probability is given by \cite{zhong2011distributed}%
\begin{equation}
P\left(  x,y,\mathbf{s}\right)  =1-\prod\nolimits_{i=1}^{N}\left[
1-p_{i}\left(  x,y,x_{i},y_{i}\right)  \right]  \text{.} \label{sq}%
\end{equation}
Finally, the coverage metric $H\left(  \mathbf{s}\right)  $ is defined as%
\[
H\left(  \mathbf{s}\right)  =\int\int_{\Omega}R\left(  x,y\right)  P\left(
x,y,\mathbf{s}\right)  dxdy.
\]
Other reasonable sensing quality metrics are also possible, as in
\cite{stipanovic2013safe} and \cite{panagou2015dynamic}. Note that $H\left(
\mathbf{s}\right)  $ is a function mapping a vector $\mathbf{s\in%
\mathbb{R}
}^{2N}$ into $%
\mathbb{R}
$.

For simplicity, in what follows we assume that all points in the mission space
are indistinguishable and set $R\left(  x,y\right)  =1$. Even though the
precise form of the function $p_{i}\left(  x,y,x_{i},y_{i}\right)  $ does not
affect our subsequent analysis, for ease of calculation in the sequel we take
it to be%
\begin{equation}
p_{i}\left(  x,y,x_{i},y_{i}\right)  =1-\frac{\left(  x-x_{i}\right)
^{2}+\left(  y-y_{i}\right)  ^{2}}{\delta_{i}^{2}}, \label{sq1}%
\end{equation}
for all $\left(  x,y\right)  \in\Omega_{i}$. 
\begin{remark}
We emphasize that the particular forms of $R\left( x,y\right) $ and $p_{i}\left(
x,y,x_{i},y_{i}\right) $ in (\ref{sq1}) are only adopted for ease of calculation. It is worth noting that
the optimal coverage theory applies to any reasonable $R\left( x,y\right) $ and $%
p_{i}\left( x,y,x_{i},y_{i}\right) $, such as%
\begin{equation*}
p_{i}\left( x,y,x_{i},y_{i}\right) =\alpha _{i}\exp \left[ -\beta _{i}\sqrt{%
\left( x-x_{i}\right) ^{2}+\left( y-y_{i}\right) ^{2}}\right]
\end{equation*}%
used in~\cite{zhong2011distributed}, where $0<\alpha _{i}\leq 1$ and $\beta
_{i}>0$ are sensing parameters.
\end{remark}

Returning to problem (\ref{of}), there are two challenges we face. First,
recall that an agent has no prior knowledge of either the mission space or the
battery levels of other agents; it only knows the location of the charging
station and of its neighbors.
In addition, the charging station is only provided with the location and battery
state information of agents when they\ are in the to-charging mode. Under this
information structure, it is clearly impossible to tackle the coverage problem
in a centralized way. The second challenge stems from the fact that, unlike
the traditional coverage problem in \cite{zhong2011distributed} where the goal
is to find the optimal equilibrium locations of agents, (\ref{of}) is a
\emph{dynamic} multi-agent coverage problem: due to the energy dynamics and
constraints in (\ref{of}), such an equilibrium may never exist, as agents move
back and forth between coverage and battery charging modes. Thus, in general,
finding the optimal speed $v_{i}^{\ast}\left(  t\right)  $ and the optimal
heading $w_{i}^{\ast}\left(  t\right)  $ in problem (\ref{of}) for all
$i=1,\ldots,N$ and all $t$ is a challenging task since its solution amounts to
a notoriously hard two-point-boundary-value problem similar to other dynamic
multi-agent optimization problems, e.g., see \cite{lin2015optimal}. In the
following, we will show how to solve this problem by modeling the combined
cooperative coverage-recharging processes as a hybrid system.

\section{HYBRID SYSTEM MODEL}

\label{hm} Our first step is to construct a hybrid system model to guarantee
that the constraints in (\ref{of}) are satisfied for all $t$. To ensure that
the problem is well-posed, we assume that%
\begin{equation}
\beta\geq N\alpha v^{2}\text{.} \label{ass}%
\end{equation}
This assumption is sufficient to guarantee the feasibility of the hybrid
system model to be constructed. In particular, by treating the charging
station as a server, the charging rate is $\beta$ if it is occupied at all
times, and referring to (\ref{bd}), the worst-case energy depletion rate over
all agents is $N\alpha v^{2}$. Thus, the condition (\ref{ass}) is sufficient
to prevent any agent from running out of energy (dying) anywhere in the
mission space. However, this assumption is not necessary in the sense that the
problem may be feasible even when (\ref{ass}) is not satisfied.

\begin{figure}[ptb]
\begin{center}
\includegraphics[width=8.4cm]{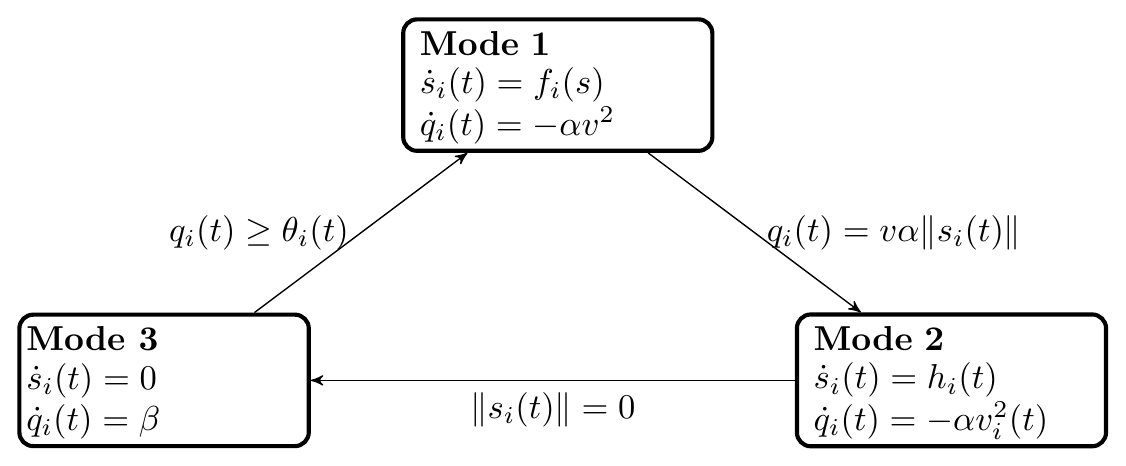}
\end{center}
\caption{A hybrid system model}%
\label{f1}%
\end{figure}

For any agent, we define three different modes: coverage (Mode 1), to-charging
(Mode 2) and in-charging (Mode 3). This hybrid system consists of a single
cycle for each agent: Mode 1$\rightarrow$Mode 2$\rightarrow$Mode
3$\rightarrow$Mode 1 as shown in Fig.~\ref{f1} and detailed next.

At Mode 1, $v_{i}\left(  t\right)  =v$ (the maximum speed for each agent), and%
\begin{align}
\cos w_{i}\left(  t\right)   &  =\frac{\frac{\partial H\left(  t\right)
}{\partial x_{i}\left(  t\right)  }}{\sqrt{\left(  \frac{\partial H\left(
t\right)  }{\partial x_{i}\left(  t\right)  }\right)  ^{2}+\left(
\frac{\partial H\left(  t\right)  }{\partial y_{i}\left(  t\right)  }\right)
^{2}}},\label{angle1}\\
\sin w_{i}\left(  t\right)   &  =\frac{\frac{\partial H\left(  t\right)
}{\partial y_{i}\left(  t\right)  }}{\sqrt{\left(  \frac{\partial H\left(
t\right)  }{\partial x_{i}\left(  t\right)  }\right)  ^{2}+\left(
\frac{\partial H\left(  t\right)  }{\partial y_{i}\left(  t\right)  }\right)
^{2}}}\text{,} \label{angle2}%
\end{align}
where the calculations of detailed expressions for $\frac{\partial H\left(
t\right)  }{\partial x_{i}\left(  t\right)  }$ and $\frac{\partial H\left(
t\right)  }{\partial y_{i}\left(  t\right)  }$ are given in Appendix~A%
. To ease notation, we rewrite the dynamics in (\ref{sd}), (\ref{sd1}) and
(\ref{bd}) as%
\begin{align}
\dot{x}_{i}\left(  t\right)   &  =f_{i}^{x}\left(  t\right)  ,\label{cx}\\
\dot{y}_{i}\left(  t\right)   &  =f_{i}^{y}\left(  t\right)  ,\label{cy}\\
\dot{q}_{i}\left(  t\right)   &  =-\alpha v^{2}.\label{cz}
\end{align}
Here $f_{i}^{x}\left(  t\right)  =v\cos w_{i}(t)$ and $f_{i}^{y}\left(
t\right)  =v\sin w_{i}(t)$, where the expressions of $\cos w_{i}(t)$ and $\sin
w_{i}(t)$ are given by (\ref{angle1}), and (\ref{angle2}), respectively.
Moreover, $f_{i}(s)$ in Fig.~\ref{f1} is given by $f_{i}(s)=[f_{i}^{x}%
,f_{i}^{y}]^{T}$. In other words, agent $i$ travels at the maximum speed, and
the heading direction follows the gradient direction of the coverage metric
with respect to agent $i$'s location. The state-of-charge of the battery
monotonically decreases with rate $\alpha v^{2}$ and when it drops to a
certain value, the agent switches to Mode 2.

A transition from Mode 1 to Mode 2 occurs when the guard function%
\begin{equation}
g_{i}\left(  s_{i},q_{i}\right)  =q_{i}\left(  t\right)  -v\alpha\left\Vert
s_{i}(t)\right\Vert \label{g12}%
\end{equation}
is zero, where $\left\Vert s_{i}(t)\right\Vert =\sqrt{x_{i}^{2}\left(
t\right)  +y_{i}^{2}\left(  t\right)  }$. At Mode 2, the speed $v_{i}\left(
t\right)  $ is determined by the scheduling algorithm used to assign an agent
to the charging station and the heading direction is constant and determined
by the location of agent $i$ at the time of switching from Mode 1 to Mode 2,
say $\tau_{2}$. Then, the motion dynamics and the state-of-charge dynamics
are:%
\begin{align}
\dot{x}_{i}\left(  t\right)   &  =-v_{i}\left(  t\right)  \frac{x_{i}\left(
\tau_{2}\right)  }{\left\Vert s_{i}(\tau_{2})\right\Vert },\label{m21}\\
\dot{y}_{i}\left(  t\right)   &  =-v_{i}\left(  t\right)  \frac{y_{i}\left(
\tau_{2}\right)  }{\left\Vert s_{i}(\tau_{2})\right\Vert },\label{m22}\\
\dot{q}_{i}\left(  t\right)   &  =-\alpha v_{i}^{2}\left(  t\right)  .
\label{m23}%
\end{align}
The speed $v_{i}\left(  t\right)  $ in Mode 2 is piecewise constant or
constant depending on which scheduling algorithm is used to resolve conflicts
when multiple agents request to use the charging station at the same time, as
discussed in Section~\ref{MR} (note that we assume no energy loss at points
where the speed may experience a jump). The function $h_{i}(t)$ in
Fig.~\ref{f1} is a column vector containing the right-hand side of (\ref{m21})
and (\ref{m22}).

A transition from Mode 2 to Mode 3 occurs when agent $i$ arrives at the
charging station and the guard function%
\begin{equation}
g_{i}\left(  s_{i}\right)  =\left\Vert s_{i}(t)\right\Vert \label{g2}
\end{equation}
is zero. At Mode 3, an agent remains at rest at the charging station,
therefore, it satisfies the dynamics%
\begin{align}
\dot{x}_{i}\left(  t\right)   &  =0, \label{m31}\\
\dot{y}_{i}\left(  t\right)   &  =0 \text{.}\label{m32}%
\end{align}
While the agent is in charging mode, the state-of-charge dynamics are given by $\dot{q}_{i}\left(  t\right)  =\beta$,
where $\beta\geq N\alpha v^{2}$ is the charging rate.

Finally, a transition from Mode 3 to Mode 1 occurs when the guard function $g_{i}\left(  q_{i}\right)  =\theta_{i}-q_{i}\left(  t\right)$
is zero, where $\theta_{i}\in(0,1]$ is a controllable threshold parameter
indicating the desired state-of-charge at which the agent may stop its
recharging process.
\begin{remark}It is worth noting that the hybrid model does not rely on a detailed energy depletion model of the state-of-charge in Mode 1. The energy consumption of sensing and communication could be included in Mode 1. When an agent switches to Mode 2, it can turn off its sensing and communication functionalities. Therefore, the energy costs in Mode 2 are only related to the agent's speed.
\end{remark}
The results on feasibility and rationality of the proposed hybrid model can be found in~\cite{meng2018hybrid}. Here we only show the results on schedulability and optimality of the parameter $\theta$.
\section{Scheduling Algorithms}\label{MR}
Since the charging station can only serve one agent at a time, a
scheduling algorithm is needed to resolve conflicts among agents competing
over access to it. Here, we consider two scheduling policies:
First-Request-First-Serve (FRFS) and Shortest-Distance-First (SDF).

\subsubsection{First Request First Serve}

Suppose that when agent $i$ sends a charging request at $\tau_{r}^{i}$, the
charging station is not reserved. Then, agent $i$ will use the maximum speed
$v$ to reach the charging station. If agent $j$ sends a charging
request at $\tau_{r}^{j}>\tau_{r}^{i}$, the arrival time of agent $j$ will be
scheduled at $\max\{\tau_{f}^{i},\tau_{a}^{j}\}$, where $\tau_{f}^{i}$ is the
time when agent $i$ finishes charging, and $\tau_{a}^{j}$ is the arrival time
if agent $j$ heads to the charging station at the maximum speed. There are two
different cases: $\tau_{f}^{i}<\tau_{a}^{j}$ and $\tau_{f}^{i}\geq\tau_{a}%
^{j}$. For the former case, there are no conflicts between agents $i$ and $j$.
This is because when agent $j$ arrives at the charging station using the
maximum speed, agent $i$ has already left the charging station. For the latter
case, the speed of agent $j$ will be set to%
\[
v_{j}\left(  t\right)  =\frac{\left\Vert s_{j}(\tau_{r}^{j})\right\Vert }%
{\tau_{f}^{i}-\tau_{r}^{j}}\leq v,
\]
for $\tau_{r}^{j}\leq t<\tau_{f}^{i}$. Therefore, agent $j$ will arrive at the
charging station right after agent $i$ finishes charging. It is
straightforward to extend the case of two agents to the case of
multiple\ competing agents.

\subsubsection{Shortest Distance First}

Suppose that agent $i$ sends a charging request at $\tau_{r}^{i}$. While agent
$i$ is on its way to the charging station, suppose that agent $j$, which is
closer to the charging station at time $\tau_{r}^{j}$, also sends a charging
request. Therefore, if both agents travel at the maximum speed, agent $j$ will
arrive at the charging station before agent $i$. In this case, the speed of
agent $j$ is set as $v_{j}\left(  t\right)  =v$,
and its arrival time is $\tau_{a}^{j}$. The arrival time of agent $i$ will be
scheduled at $\max\{\tau^{j}_{f},\tau^{i}_{a}\}$, where $\tau_{f}^{j}$ is the
leaving time of agent $j$ from the charging station and $\tau_{a}^{i}$ is the
intended arrival time of agent $i$ to the charging station. Similarly, there
are two different cases: $\tau_{f}^{j}<\tau_{a}^{i}$, and $\tau_{f}^{j}%
\geq\tau_{a}^{i}$. For the former case, there are no conflicts between agents
$j$ and $i$. For the latter case, the speed of agent $i$ is set as%
\[
v_{i}\left(  t\right)  =\left\{
\begin{array}
[c]{cc}%
v & \text{for }t\in\lbrack\tau_{r}^{i},\tau_{r}^{j}),\\
\frac{\left\Vert s_{i}(\tau_{r}^{j})\right\Vert }{\tau_{f}^{j}-\tau_{r}^{j}} &
\text{for }t\in\lbrack\tau_{r}^{j},\tau_{f}^{j}).
\end{array}
\right.
\]
In this case, agent $i$ is scheduled to arrive at the charging station right
after agent $j$ finishes charging. It is not difficult to extend this
reasoning to the case of multiple agents: the one closer to the charging
station always receives the highest priority to be served first.

\section{Main Results}\label{IPAC}
We now address the question of selecting an optimal charging level, denoted by
$\mathbf{\theta}=\left[  \theta_{1},\ldots,\theta_{N}\right]  $, when an agent
is in the charging mode. This problem boils down to optimizing the parameter
$\mathbf{\theta}$ so that the objective function in (\ref{of}) is maximized.
By writing explicitly the dependence on $\mathbf{\theta}$, the optimization
problem becomes%
\[
J\left(  \mathbf{\theta}\right)  =\max_{\mathbf{\theta}}\frac{1}{T}\int
_{0}^{T}H\left(  s\left(  \mathbf{\theta},t\right)  \right)  dt\text{.}%
\]

Even though $\mathbf{\theta}$ is only used in Mode 3, its optimal value
affects the entire hybrid system model. By controlling $\mathbf{\theta}$, we
directly control the switching times of agents from Mode 3 to Mode 1, and
indirectly control the switching times of agents from Mode 1 to Mode 2. The
switching times of agents from Mode 2 to Mode 3 are controlled by the proposed
scheduling algorithms. Also note that the parameter $\mathbf{\theta}$ is
constant. We can obtain optimal charging thresholds through off-line analysis
and implement the coverage task on line by all agents in distributed fashion.
We use IPA~\cite{cassandras2010perturbation} to determine the optimal $\mathbf{\theta}$.

Before proceeding, we briefly review the IPA framework for general stochastic hybrid systems
as presented in \cite{cassandras2010perturbation}, which plays an instrumental role in obtaining the optimal dwell time of all agents at the charging station.

Let $\{\tau_{k}(\theta)\}$, $k=1,\ldots,K$, denote the occurrence times of all
events in the state trajectory of a hybrid system with dynamics $\dot
{x}\ =\ f_{k}(x,\theta,t)$ over an interval $[\tau_{k}(\theta),\tau
_{k+1}(\theta))$, where $\theta\in\Theta$ is some parameter vector and
$\Theta$ is a given compact, convex set. For convenience, we set $\tau_{0}=0$
and $\tau_{K+1}=T$. We use the Jacobian matrix notation: $x^{\prime}%
(t)\equiv\frac{\partial x(\theta,t)}{\partial\theta}$ and $\tau_{k}^{\prime
}\equiv\frac{\partial\tau_{k}(\theta)}{\partial\theta}$, for all state and
event time derivatives. It is shown in \cite{cassandras2010perturbation} that
\begin{equation}
\frac{d}{dt}x^{\prime}(t)=\frac{\partial f_{k}(t)}{\partial x}x^{\prime
}(t)+\frac{\partial f_{k}(t)}{\partial\theta}, \label{eq:IPA_1}%
\end{equation}
for $t\in\lbrack\tau_{k},\tau_{k+1})$ with boundary condition:
\begin{equation}
x^{\prime}(\tau_{k}^{+})=x^{\prime}(\tau_{k}^{-})+[f_{k-1}(\tau_{k}^{-}%
)-f_{k}(\tau_{k}^{+})]\tau_{k}^{\prime} , \label{eq:IPA_2}%
\end{equation}
for $k=0,...,K$. In order to complete the evaluation of $x^{\prime}(\tau
_{k}^{+})$ in (\ref{eq:IPA_2}), we need to determine $\tau_{k}^{\prime}$. We
classify events into two categories. An event is exogenous if it causes a
discrete state transition at time $\tau_{k}$ independent of the controllable
vector $\theta$ and, therefore, satisfies $\tau_{k}^{\prime}=0$. Otherwise,
the event is \emph{endogenous} and there exists a continuously differentiable
function $g_{k}:\mathbb{R}^{n}\times\Theta\rightarrow\mathbb{R}$ such that
$\tau_{k}\ =\ \min\{t>\tau_{k-1}\ :\ g_{k}\left(  x\left(  \theta,t\right)
,\theta\right)  =0\}$ and
\begin{equation}
\tau_{k}^{\prime}=-[\frac{\partial g_{k}}{\partial x}f_{k}(\tau_{k}^{-}%
)]^{-1}(\frac{\partial g_{k}}{\partial\theta}+\frac{\partial g_{k}}{\partial
x} x^{\prime}(\tau_{k}^{-})) \label{eq:IPA_3}%
\end{equation}
as long as $\frac{\partial g_{k}}{\partial x}f_{k}(\tau_{k}^{-})\neq0$
(details may be found in \cite{cassandras2010perturbation}).

Denote the time-varying cost along a given trajectory as $L(x,\theta,t)$, so
the cost in the $k$-th inter-event interval is $J_{k}(x,\theta)=\int_{\tau
_{k}}^{\tau_{k+1}}L(x,\theta,t)dt$ and the total cost is $J(x,\theta
)=\sum_{k=0}^{K}J_{k}(x,\theta)$. Differentiating and applying the Leibniz
rule with the observation that all terms of the form $L(x(\tau_{k}%
),\theta,\tau_{k})\tau_{k}^{\prime}$ are mutually canceled with $\tau
_{0}=0,\tau_{K+1}=T$ fixed, we obtain%
\begin{align}
\frac{\partial J(x,\theta)}{\partial\theta} &  =\sum_{k=0}^{K}\frac{\partial
}{\partial\theta}\int_{\tau_{k}}^{\tau_{k+1}}L(x,\theta,t)dt\nonumber\\
&  =\sum_{k=0}^{K}\int_{\tau_{k}}^{\tau_{k+1}}\frac{\partial L(x,\theta
,t)}{\partial x}x^{\prime}(t)+\frac{\partial L(x,\theta,t)}{\partial\theta
}dt.\label{eq:GeneralGradientParametricObj}%
\end{align}

Now let us return to our problem and define the following notations%
\[
\tau^{\prime}_{k}=\frac{\partial \tau_{k}(\theta)}{\partial \theta},\quad \mathbf{x}^{\prime}_{i}=\frac{\partial x_{i}(\theta)}{\partial \theta},\quad \mathbf{y}^{\prime}_{i}=\frac{\partial y_{i}(\theta)}{\partial \theta}
\]%
which are row vectors, and%
\begin{align*}
\frac{\partial\mathbf{x}\left(\theta,  t\right)  }{\partial\mathbf{\theta}} &
=\left[  \mathbf{x}_{1}^{\prime}\left( \theta, t\right)^{T}  ,\cdots,\mathbf{x}%
_{N}^{\prime}\left( \theta, t\right)^{T}  \right]  ^{T}\text{,}\\
\frac{\partial\mathbf{y}\left(  \theta,t\right)  }{\partial\mathbf{\theta}}  &
=\left[  \mathbf{y}_{1}^{\prime}\left(  \theta,t\right)^{T}   ,\cdots,\mathbf{y}%
_{N}^{\prime}\left( \theta, t\right)^{T}   \right]  ^{T}\text{,}
\end{align*}%
are matrices.

Let us assume that all agents start with the battery level%
\[
q_{i}\left(  0\right)  >v\alpha\left\Vert s_{i}\left(  0\right)  \right\Vert
\text{,}%
\]
for $i=1,\ldots,N,$ that is, all agents start with Mode 1.

For $t\in\left[  \tau_{1},\tau_{2}\right)$, applying (\ref{eq:IPA_1}) to (\ref{cx}) and (\ref{cy}) yields that%
\begin{align}
\frac{d}{dt}\mathbf{x}_{i}^{\prime}\left( \theta, t\right)    & =v\frac{\partial\cos
w_{i}\left(  t\right)  }{\partial\mathbf{x}\left(  \theta,t\right)  }\frac
{\partial\mathbf{x}\left( \theta, t\right)  }{\partial\mathbf{\theta}}+v\frac
{\partial\cos w_{i}\left(  t\right)  }{\partial\mathbf{y}\left(\theta,  t\right)
}\frac{\partial\mathbf{y}\left( \theta, t\right)  }{\partial\mathbf{\theta}%
},\label{D1}\\
\frac{d}{dt}\mathbf{y}_{i}^{\prime}\left(\theta,  t\right)    & =v\frac{\partial\sin
w_{i}\left(  t\right)  }{\partial\mathbf{x}\left( \theta, t\right)  }\frac
{\partial\mathbf{x}\left( \theta, t\right)  }{\partial\mathbf{\theta}}+v\frac
{\partial\sin w_{i}\left(  t\right)  }{\partial\mathbf{y}\left( \theta, t\right)
}\frac{\partial\mathbf{y}\left( \theta, t\right)  }{\partial\mathbf{\theta}%
},\label{D2}%
\end{align}
where the detailed calculations of $\frac{\partial\cos
w_{i}\left(  t\right)  }{\partial\mathbf{x}\left(  \theta,t\right)  }$, $\frac
{\partial\cos w_{i}\left(  t\right)  }{\partial\mathbf{y}\left(\theta,  t\right)
}$, $\frac{\partial\sin
w_{i}\left(  t\right)  }{\partial\mathbf{x}\left( \theta, t\right)  }$, and $\frac
{\partial\sin w_{i}\left(  t\right)  }{\partial\mathbf{y}\left( \theta, t\right)
}$ can be founded in Appendix~B.
Note that for agents $j\notin \mathcal{N}_{i}$,%
\[
\frac{\partial\cos
w_{i}\left(  t\right)  }{\partial x_{j}\left(  t\right)  }=0, \quad \frac{\partial\sin
w_{i}\left(  t\right)  }{\partial x_{j}\left(  t\right)  }=0.
\]

For the state-of-charge, we have%
\[
\frac{d}{dt}\mathbf{q}_{i}^{\prime}\left(\theta,  t\right)  =\mathbf{0},
\]
by applying (\ref{eq:IPA_1}) to (\ref{cz}), which implies that $\mathbf{q}_{i}^{\prime}\left( \theta, \tau_{2}^{-}\right)  =\mathbf{q}_{i}^{\prime
}\left( \theta, \tau_{1}^{+}\right)$.
By solving the
differential equations (\ref{D1}) and (\ref{D2}), we can obtain $\mathbf{x}_{i}^{\prime}\left(\theta,  \tau_{2}^{-}\right)  $ and
$\mathbf{y}_{i}^{\prime}\left( \theta, \tau_{2}^{-}\right)  $.

At $\tau_{2},$ the guard condition%
\begin{align*}
&g_{i}\left(  x_{i}\left(  \theta,\tau
_{2}\right)  ,y_{i}\left(  \theta,\tau_{2}\right)  ,q_{i}\left(  \theta
,\tau_{2}\right)  \right) \\ =&q_{i}^{2}\left(  \theta,\tau_{2}\right)
-v^{2}\alpha^{2}\|s_{i}\left(  \theta,\tau_{2}\right)\| ^{2} =0.
\end{align*}%
Note that we use an equivalent form of (\ref{g12}) by squaring both terms for an easier calculation of derivatives.
This is an endogenous event. By applying (\ref{eq:IPA_3}) to the above guard function and the dynamics in (\ref{cx}), (\ref{cy}) and (\ref{cz}), we have%
{\small\[
\mathbf{\tau}_{2}^{\prime}=\frac{q_{i}\left(  \tau_{2}\right)  \mathbf{q}%
_{i}^{\prime}\left(  \tau_{2}^{-}\right)  -v^{2}\alpha^{2}\left[  x_{i}\left(
\tau_{2}\right)  \mathbf{x}_{i}^{\prime}\left(  \tau_{2}^{-}\right)
+y_{i}\left(  \tau_{2}\right)  \mathbf{y}_{i}^{\prime}\left(  \tau_{2}%
^{-}\right)  \right]  }{\alpha v^{2}q_{i}\left(  \tau_{2}\right)  +v^{3}%
\alpha^{2}\left[  x_{i}\left(  \tau_{2}\right)  \cos w_{i}\left(  \tau_{2}%
^{-}\right)  +y_{i}\left(  \tau_{2}\right)  \sin w_{i}\left(  \tau_{2}%
^{-}\right)  \right]  }
\]}
with the boundary conditions%
\begin{align*}
\mathbf{q}_{i}^{\prime}\left(  \tau_{2}^{+}\right)  &=\mathbf{q}_{i}^{\prime
}\left(  \tau_{2}^{-}\right)  +\alpha\left[  v_{i}^{2}\left(  \tau_{2}^{+}\right)
-v^{2}\right]  \mathbf{\tau}_{2}^{\prime}\\
\mathbf{x}_{i}^{\prime}\left(  \tau_{2}^{+}\right)   &  =\mathbf{x}%
_{i}^{\prime}\left(  \tau_{2}^{-}\right)  +\left[  v\cos w_{i}\left(  \tau
_{2}^{-}\right)  -v_{i}\left(  \tau_{2}^{+}\right)  \cos w_{i}\left(  \tau
_{2}^{+}\right)  \right]  \mathbf{\tau}_{2}^{\prime},\\
\mathbf{y}_{i}^{\prime}\left(  \tau_{2}^{+}\right)   &  =\mathbf{y}%
_{i}^{\prime}\left(  \tau_{2}^{-}\right)  +\left[  v\sin w_{i}\left(  \tau
_{2}^{-}\right)  -v_{i}\left(  \tau_{2}^{+}\right)  \sin w_{i}\left(  \tau
_{2}^{+}\right)  \right]  \mathbf{\tau}_{2}^{\prime},
\end{align*}
which are obtained by applying (\ref{eq:IPA_2}) to the dynamics in (\ref{m21}), (\ref{m22}) and (\ref{m23}).
\begin{remark}Irrespective of the scheduling algorithm, if agent $i$ is the first to request charging in the current queue, then $v_{i}\left(  \tau_{2}^{+}\right) =v$, and $\textbf{q}_{i}^{\prime}\left( \theta, \tau_{2}^{+}\right)  =\textbf{q}_{i}^{\prime}\left(\theta , \tau_{2}%
^{-}\right)=\textbf{q}_{i}^{\prime}\left(\theta,  \tau_{1}^{+}\right)$.
\end{remark}

In Mode 2, the right-hand sides of (\ref{m21}), (\ref{m22}), and
(\ref{m23}) are constant or piecewise constant depending on the scheduling algorithm. Therefore, we have%
\begin{align*}
&\frac{d}{dt}\mathbf{x}_{i}^{\prime}\left(\theta,  t\right)=0, \quad \frac{d}{dt}\mathbf{y}_{i}^{\prime}\left(\theta,  t\right)  =0,\\
&\frac{d}{dt}\mathbf{q}_{i}^{\prime}\left(\theta,  t\right) =0,
\end{align*}%
according to (\ref{eq:IPA_1}).
It is easy to see that%
\begin{align}
\mathbf{x}_{i}^{\prime}\left( \theta, \tau_{3}^{-}\right)   &  =\mathbf{x}%
_{i}^{\prime}\left( \theta, \tau_{2}^{+}\right)  ,\label{IPA1}\\
\mathbf{y}_{i}^{\prime}\left( \theta, \tau_{3}^{-}\right)   &  =\mathbf{y}%
_{i}^{\prime}\left( \theta, \tau_{2}^{+}\right)  ,\label{IPA2}\\
\mathbf{q}_{i}^{\prime}\left( \theta, \tau_{3}^{-}\right)   &  =\mathbf{q}%
_{i}^{\prime}\left( \theta, \tau_{2}^{+}\right)  . \label{IPA3}%
\end{align}

In the SDF scheduling algorithm, the velocity of agent $i$ may be adjusted due to the competition to the charging station. This is the case when agent $j$, which is closer to the
charing station than agent $i$, requests for the charging service. Such events are independent of $\theta$ and are, therefore, treated as exogenous events. In Mode 2, the relationships (\ref{IPA1}), (\ref{IPA2}), and (\ref{IPA3}) hold
independent of the scheduling methods, and the number of exogenous events.

At time $\tau_{3}$, the guard function $g_{i}\left(  x_{i}\left(  \theta
,\tau_{3}\right)  ,y_{i}\left(  \theta,\tau_{3}\right)  \right)  =\|s_{i}%
\left(  \tau_{3}\right)\|^{2}  =0$. Again, an equivalent form of the guard function (\ref{g2}) is used by squaring the term for an easier calculation of derivatives. This is an endogenous event. According to (\ref{eq:IPA_3}), we can
calculate%
{\small\[
\mathbf{\tau}_{3}^{\prime}=-\frac{x_{i}\left( \theta, \tau_{3}\right)  \mathbf{x}%
_{i}^{\prime}\left(\theta,  \tau_{3}^{-}\right)  +y_{i}\left(\theta,  \tau_{3}\right)
\mathbf{y}_{i}^{\prime}\left( \theta, \tau_{3}^{-}\right)  }{x_{i}\left( \theta, \tau
_{3}\right)  v_{i}\left( \tau_{3}^{-}\right)  \cos w_{i}\left(  \tau_{3}%
^{-}\right)  +y_{i}\left( \theta, \tau_{3}\right)  v_{i}\left(  \tau_{3}^{-}\right)
\sin w_{i}\left(  \tau_{3}^{-}\right)  }%
\]}%
based on the dynamics (\ref{m21}), (\ref{m22}) and (\ref{m23})
and the boundary conditions are%
\begin{align*}
\mathbf{x}_{i}^{\prime}\left( \theta, \tau_{3}^{+}\right)    & =\mathbf{x}%
_{i}^{\prime}\left( \theta, \tau_{3}^{-}\right)  +v_{i}\left(  \tau_{3}^{-}\right)
\cos w_{i}\left(  \tau_{3}^{-}\right)  \mathbf{\tau}_{3}^{\prime}\\
\mathbf{y}_{i}^{\prime}\left( \theta, \tau_{3}^{+}\right)    & =\mathbf{y}%
_{i}^{\prime}\left( \theta, \tau_{3}^{-}\right)  +v_{i}\left(  \tau_{3}^{-}\right)
\sin w_{i}\left(  \tau_{3}^{-}\right)  \mathbf{\tau}_{3}^{\prime}\\
\mathbf{q}_{i}^{\prime}\left(  \tau_{3}^{+}\right)    & =\mathbf{q}%
_{i}^{\prime}\left( \theta, \tau_{3}^{-}\right)  -\left[  \alpha v_{i}^{2}\left(
\tau_{3}^{-}\right)  +\beta\right]  \mathbf{\tau}_{3}^{\prime}%
\end{align*}
by applying (\ref{eq:IPA_2}) to the dynamics in (\ref{m31}), (\ref{m32}) and (\ref{b_charge}).
\begin{remark}
When calculating $\mathbf{\tau}_{3}^{\prime}$, we find that both the numerator and denominator are zero due to $x_{i}\left( \theta, \tau_{3}\right)=y_{i}\left( \theta, \tau_{3}\right)=0$. In this case, the value of $\mathbf{\tau}_{3}^{\prime}$ is calculated according to its limit in the direction $w_{i}(\mathbf{\tau}_{3}^{-})$. Let us put $x_{i}$ and $y_{i}$ in the polar coordinate, then $x_{i}=r\cos w_{i}\left(  \tau_{3}^{-}\right)$ and $y_{i}=r\sin w_{i}\left(  \tau_{3}^{-}\right)$. Replacing $x_{i}$ and $y_{i}$ in $\mathbf{\tau}_{3}^{\prime}$, it becomes%
\begin{align*}
\mathbf{\tau }_{3}^{\prime } &=-\lim_{r\rightarrow 0}\frac{r\cos w_{i}\left(
\tau _{3}^{-}\right) \mathbf{x}_{i}^{\prime }\left( \theta ,\tau
_{3}^{-}\right) +r\sin w_{i}\left( \tau _{3}^{-}\right) \mathbf{y}%
_{i}^{\prime }\left( \theta ,\tau _{3}^{-}\right) }{rv_{i}\left( \tau
_{3}^{-}\right) \cos w_{i}^{2}\left( \tau _{3}^{-}\right) +rv_{i}\left( \tau
_{3}^{-}\right) \sin w_{i}^{2}\left( \tau _{3}^{-}\right) } \\
&=-\frac{\cos w_{i}\left( \tau _{3}^{-}\right) \mathbf{x}_{i}^{\prime
}\left( \theta ,\tau _{3}^{-}\right) +\sin w_{i}\left( \tau _{3}^{-}\right)
\mathbf{y}_{i}^{\prime }\left( \theta ,\tau _{3}^{-}\right) }{v_{i}\left(
\tau _{3}^{-}\right) }.
\end{align*}
\end{remark}

Note that in Mode 2, agents do not change their direction and $w_{i}\left(
\tau_{3}^{-}\right)  =w_{i}\left(  \tau_{2}^{+}\right)  $.

In Mode 3, during the cycle $\left[  \tau_{3},\tau_{1}\right)  ,$ we can
obtain%
\begin{align*}
\frac{d}{dt}\mathbf{x}_{i}^{\prime}\left( \theta, t\right)   =0, \quad \frac{d}{dt}\mathbf{y}_{i}^{\prime}\left( \theta, t\right)   =0,\quad \frac{d}{dt}\mathbf{q}_{i}^{\prime}\left( \theta, t\right)    =0
\end{align*}
by applying (\ref{eq:IPA_1}) to the dynamic equations (\ref{m31}), (\ref{m32}) and (\ref{b_charge}).
Therefore, it is easy to calculate%
\begin{align*}
\mathbf{x}_{i}^{\prime}\left( \theta, \tau_{1}^{-}\right)  & =\mathbf{x}%
_{i}^{\prime}\left( \theta, \tau_{3}^{+}\right), \quad \mathbf{y}_{i}^{\prime}\left(\theta,  \tau_{1}^{-}\right) =\mathbf{y}%
_{i}^{\prime}\left( \theta, \tau_{3}^{+}\right)  \\
\mathbf{q}_{i}^{\prime}\left( \theta, \tau_{1}^{-}\right)    & =\mathbf{q}%
_{i}^{\prime}\left( \theta, \tau_{3}^{+}\right)  \text{.}%
\end{align*}

At time $\tau_{1}$, the threshold%
\[
g_{i}\left(  q_{i}\left(  \theta,\tau
_{1}\right)  \right)  =q_{i}\left(  \theta,\tau_{1}\right)  -\theta_{i}=0.
\]
This is an endogenous event. We can obtain%
\[
\mathbf{\tau}_{1}^{\prime}=\frac{1-\textbf{q}_{i}^{\prime}\left(  \theta,\tau_{1}^{-}\right)}{\beta},
\]
and the boundary conditions%
\begin{align*}
\mathbf{x}_{i}^{\prime}\left(\theta,  \tau_{1}^{+}\right)    & =\mathbf{x}%
_{i}^{\prime}\left( \theta, \tau_{1}^{-}\right)  -v\cos w_{i}\left(  \tau_{1}%
^{+}\right)  \mathbf{\tau}_{1}^{\prime}\\
\mathbf{y}_{i}^{\prime}\left( \theta, \tau_{1}^{+}\right)    & =\mathbf{y}%
_{i}^{\prime}\left(  \theta,\tau_{1}^{-}\right)  -v\sin w_{i}\left(  \tau_{1}%
^{+}\right)  \mathbf{\tau}_{1}^{\prime}\\
\mathbf{q}_{i}^{\prime}\left( \theta, \tau_{1}^{+}\right)    & =\mathbf{q}%
_{i}^{\prime}\left( \theta, \tau_{1}^{-}\right)  +\left(  \beta+\alpha v^{2}\right)
\mathbf{\tau}_{1}^{\prime}\text{,}%
\end{align*}
according to (\ref{eq:IPA_3}) and (\ref{eq:IPA_2}), respectively, based on the dynamics in (\ref{m31}), (\ref{m32}), (\ref{b_charge}), (\ref{cx}), (\ref{cy}) and (\ref{cz}).
Now the IPA derivative of $dJ/d\theta$ can be obtained by taking derivatives of $J(\theta)$ with respective to $\theta$ as shown in~(\ref{eq:GeneralGradientParametricObj}):%
\[
\frac{dJ}{d\theta }=\sum_{k=0}^{l}\frac{d}{d\theta }%
\int_{t_{k}}^{t_{k+1}}H_{k}\left( s,\theta ,t\right) dt
\]%
and applying the Leibnitz rule we obtain, for every $k=0,\ldots ,l,$%
\begin{eqnarray*}
&&\frac{d}{d\theta }\int_{t_{k}}^{t_{k+1}}H_{k}\left( s,\theta ,t\right) dt
\\
&=&\int_{t_{k}}^{t_{k+1}}\left[ \frac{\partial H_{k}\left( s,\theta
,t\right) }{\partial \mathbf{x}}\mathbf{x}^{\prime }+\frac{\partial
H_{k}\left( s,\theta ,t\right) }{\partial \mathbf{y}}\mathbf{y}^{\prime }%
\right] dt \\
&&+H_{k}\left( s\left( t_{k+1}\right) ,\theta ,t_{k+1}\right) \mathbf{t}%
_{k+1}^{\prime }-H_{k}\left( s\left( t_{k}\right) ,\theta ,t_{k}\right)
\mathbf{t}_{k}^{\prime }
\end{eqnarray*}%
where $t_{k}$ are event times of any agents, $t_{0}=0$ and $t_{l}=T$.
The parameter $\theta$ is updated as%
\begin{equation}
\theta_{n+1}=\theta_{n}+\lambda_{n} \frac{dJ(\theta_{n})}{d\theta_{n} },\label{thetaupdating}
\end{equation}%
where $\{\lambda_{n}\}$ is a step size sequence.

\section{SIMULATION RESULTS}\label{SIM}
In this section, we illustrate the optimization process in (\ref{thetaupdating}), and compare the performance by using different scheduling algorithms (FRFS, and SDF).

\begin{figure}[ptb]
\begin{center}
\includegraphics[scale=0.5]{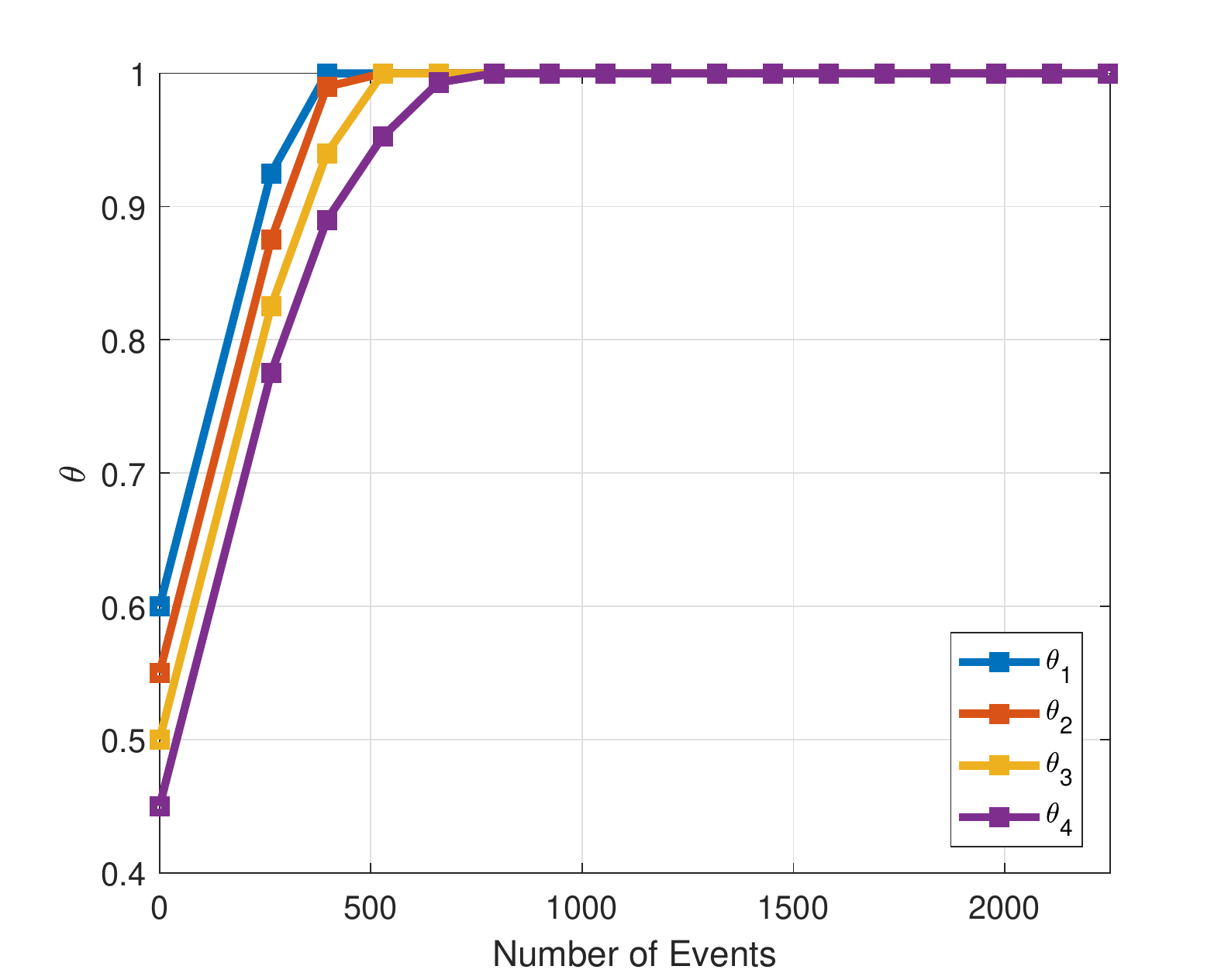}
\end{center}
\caption{The evolution of $\theta$ under the FRFS scheduling method}%
\label{f2}%
\vspace{-0.5cm}
\end{figure}

\begin{figure}[ptb]
\begin{center}
\includegraphics[scale=0.5]{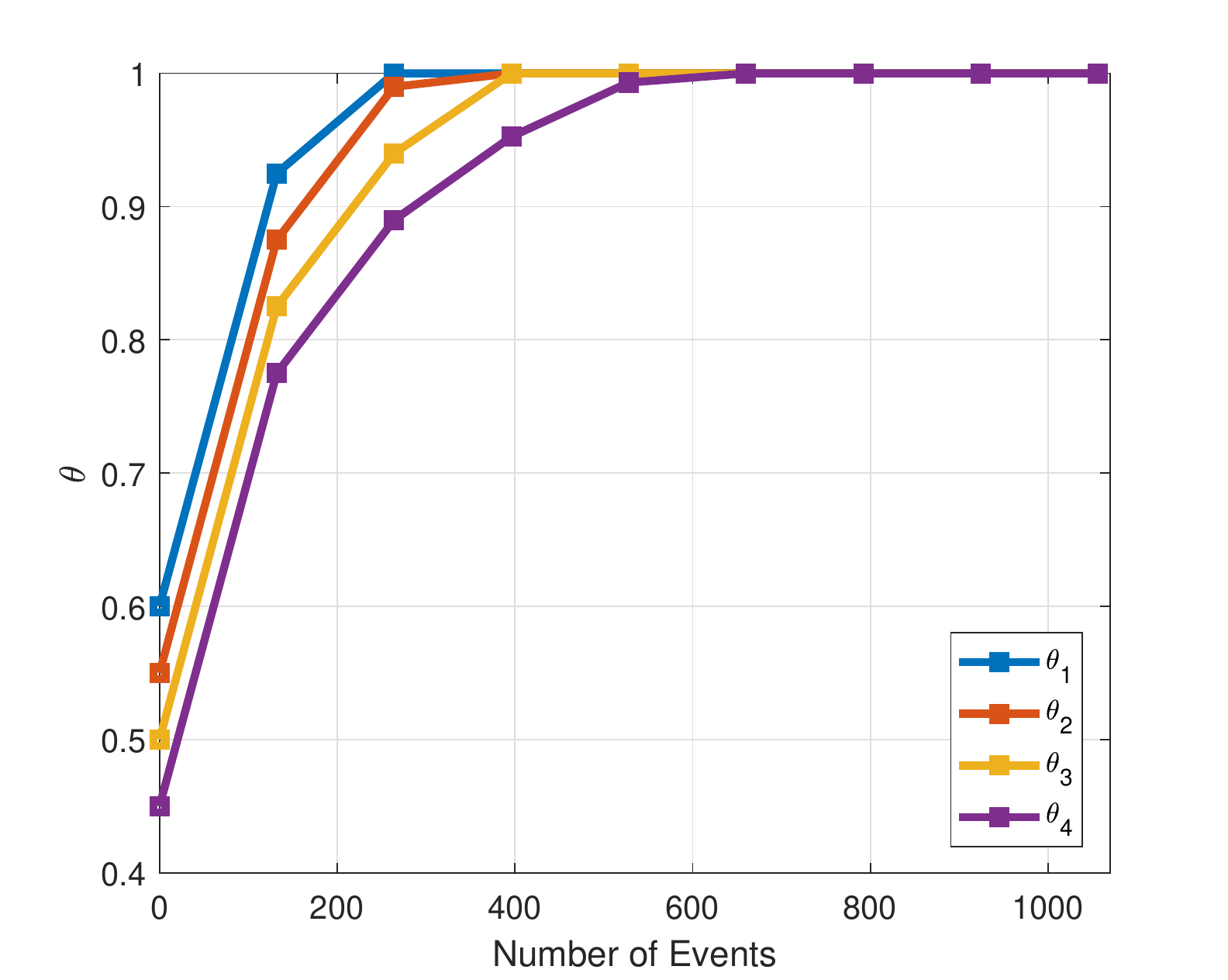}
\end{center}
\caption{The evolution of $\theta$ under the SDF scheduling method}%
\label{f3}%
\vspace{-0.5cm}
\end{figure}

\begin{figure}[ptb]
\begin{center}
\includegraphics[scale=0.5]{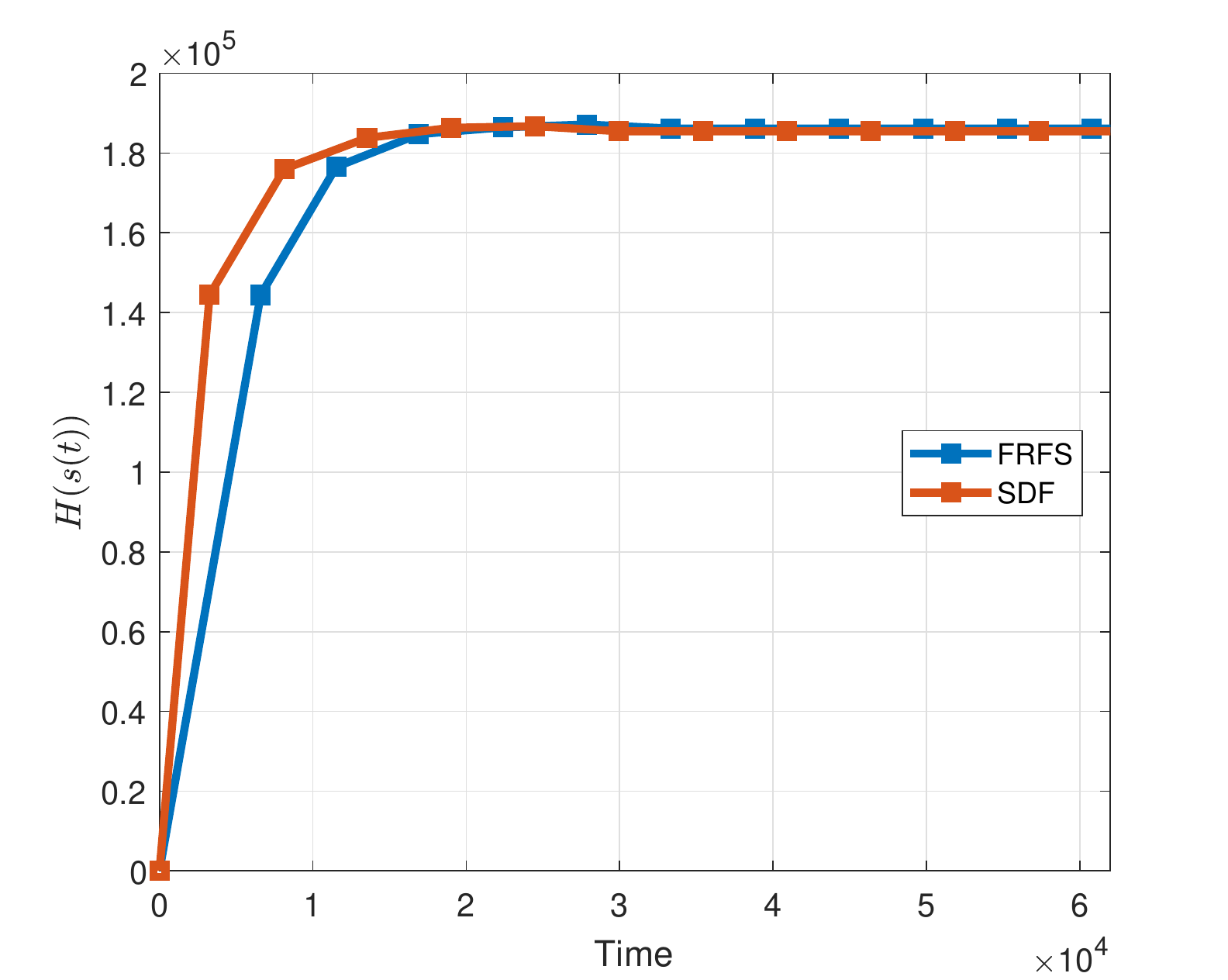}
\end{center}
\caption{The comparison of performance using different scheduling algorithms}%
\label{f4}%
\vspace{-0.5cm}
\end{figure}

The mission space is a 60 by 50 rectangular area without obstacles. We consider a team of four agents with initial locations (2,2), (4,4), (6,6) and (8,8). The initial state-of-charge variables are randomly generated, which are $97\%$, $48\%$, $71\%$, and $46\%$, respectively. The maximum speed is $v=5$, and the sensing range $\delta_{i}=22$ for all $i=1,\ldots,4$. The parameter $\alpha=0.0001$, and $\beta=4\alpha v^{2}=0.01$.
Figures~\ref{f2} and~\ref{f3} show the evolution of $\theta$ under the FRFS and SDF scheduling algorithms, respectively, where the step size sequence $\{\lambda_{n}\}$ over iterations $n=0,1,\ldots$, is chosen as $\{(\|\frac{dJ(\theta_{n})}{d\theta_{n}}\|n^{\frac{3}{2}})^{-1}\}$.
 It can be seen from both figures that it is optimal to fully charge the battery for both scheduling algorithms. The simulation runs for $T=5400$ by considering the optimal $\theta=1$. The comparison of the coverage performance between different scheduling algorithms is depicted in Fig.~\ref{f4}. The coverage performance is $J(\theta)=186407$ for FRFS, and $J(\theta)=186095$ for SDF, respectively. The difference between the performance of the two scheduling algorithms is within $0.1\%$. Therefore, no general conclusions can be drawn on which scheduling algorithm is better, even though we might expect SDF to be preferable because it uses the distance information compared to FRFS.

A visual interactive simulation can be found at
\url{http://www.bu.edu/codes/simulations/Coverage_ADHS}. Interested readers
are encouraged to interact with the simulation by choosing different
scheduling algorithms, as well as adjusting parameters such as the number of
agents $N$, the sensing range $\delta_{i}$, or the maximum speed $v$.

\section{CONCLUSIONS}\label{con}
A hybrid system model is proposed to capture the behavior of
multiple agents cooperatively solving an optimal coverage problem under energy
depletion and repletion constraints. The proposed model links each agent's
coverage, to-charging, and in-charging modes so as to form a cycle and the
guard conditions are designed to maximize the coverage performance over a
finite time horizon as well as to ensure that the agents never run out of
energy. Full repletion is optimal to
maximize the coverage objective function as shown by numerical calculations using IPA; and a theoretical proof is the subject of ongoing research.
We are also working on the inclusion of energy expended for communication among agents (see
Remark 1). Finally,
when obstacles are present in the mission space, finding optimal trajectories
for agents in Mode 2 is a challenging task that we plan to address in future
work.




\section*{APPENDIX}

\subsection{Calculation of the Gradient}\label{apa}
To find the heading direction of agent $i$, we need to calculate
the gradient of $H\left(  \mathbf{s}\right)  $ at point $(x_{i},y_{i})$, which
is%
\[
\nabla H\left(  s_{i}\right)  =\left[  \frac{\partial H}{\partial x_{i}}%
,\frac{\partial H}{\partial y_{i}}\right]  ^{T} \text{.}%
\]
According to~\cite{flanders1973differentiation}, we can calculate the gradient
as%
\[
\frac{\partial H}{\partial x_{i}}=\iint\nolimits_{\Omega}\frac{\partial
P}{\partial x_{i}}+\int_{\partial\Omega}P\left(  \frac{\partial x}{\partial
x_{i}}dy-\frac{\partial y}{\partial x_{i}}dx\right)  ,
\]
where the integration in the second term is done in the counterclockwise
direction over the boundary of $\Omega$. Recalling the expressions of
(\ref{sq}) and (\ref{sq1}), we have%
\[
\int_{\partial\Omega}P\left(  \frac{\partial x}{\partial x_{i}}dy-\frac
{\partial y}{\partial x_{i}}dx\right)  =0\text{.}%
\]
This is because when $\left(  x,y\right)  \in\partial\Omega\cap\partial
\Omega_{i}$, $P=0$; when $\left(  x,y\right)  \in\partial\Omega\backslash
\partial\Omega_{i}$,%
\[
\frac{\partial x}{\partial x_{i}}=\frac{\partial y}{\partial x_{i}}=0\text{.}%
\]
Therefore, we can obtain%
\begin{equation}
\frac{\partial H}{\partial x_{i}}=\iint\limits_{\Omega_{i}}\frac{2\left(
x-x_{i}\right)  }{\delta_{i}^{2}}\prod\limits_{j\in\mathcal{N}_{i}}\left(
1-p_{j}\right)  dxdy. \label{cal1}%
\end{equation}
Similarly, we have%
\begin{equation}
\frac{\partial H}{\partial y_{i}}=\iint\limits_{\Omega_{i}}\frac{2\left(
y-y_{i}\right)  }{\delta_{i}^{2}}\prod\limits_{j\in\mathcal{N}_{i}}\left(
1-p_{j}\right)  dxdy. \label{cal2}%
\end{equation}

\begin{remark}
When the sensing range $\Omega_{i}$ of agent $i$ is blocked by the boundary, the gradient can be derived similarly using a simple projection onto the feasible mission space.
The detailed calculations for this case are thus not shown here.
\end{remark}

\subsection{Derivative of the Gradient}

Here we discuss the calculations of $\frac{\partial\cos w_{i}}{\partial\mathbf{x}%
}$, and the formulas for $\frac{\partial\cos w_{i}}{\partial\mathbf{y}}$,
$\frac{\partial\sin w_{i}}{\partial\mathbf{x}}$, $\frac{\partial\sin w_{i}%
}{\partial\mathbf{y}}$ can be derived similarly.

Recalling the definition of $\cos w_{i}$ in (\ref{angle1}), we take the derivative of $\cos w_{i}$ with respect to $x_{i}$, and have%
\begin{align*}
\frac{\partial\cos w_{i}}{\partial x_{i}} &  =\frac{\left\Vert \nabla H\left(
s_{i}\right)  \right\Vert \frac{\partial^{2}H}{\partial x_{i}^{2}}%
-\frac{\partial H}{\partial x_{i}}\frac{\frac{\partial H}{\partial x_{i}}\frac{\partial^{2}H}{\partial
x_{i}^{2}}+\frac{\partial H}{\partial y_{i}}\frac{\partial^{2}H}{\partial x_{i}\partial y_{i}}}{\left\Vert
\nabla H\left(  s_{i}\right)  \right\Vert }}{\left\Vert \nabla H\left(
s_{i}\right)  \right\Vert ^{2}}\\
&  =\frac{\frac{\partial^{2}H}{\partial x_{i}^{2}}}{\left\Vert \nabla H\left(
s_{i}\right)  \right\Vert }-\frac{\cos w_{i}\left(  t\right) \left(  \frac{\partial H}{\partial x_{i}}\frac{\partial^{2}H}{\partial
x_{i}^{2}}+\frac{\partial H}{\partial y_{i}}\frac{\partial^{2}H}{\partial x_{i}\partial y_{i}}\right) }{\left\Vert
\nabla H\left(  s_{i}\right)  \right\Vert ^{2}}.
\end{align*}
In order to obtain $\frac{\partial\cos w_{i}}{\partial\mathbf{x}},$ we need to
calculate $\frac{\partial^{2}H}{\partial x_{i}^{2}}$ and $\frac{\partial^{2}%
H}{\partial y_{i}\partial x_{i}}$, which will be given as follows. Taking the
partial derivative of (\ref{cal1}) with respect to $x_{i}$, we have%
\begin{align*}
\frac{\partial^{2}H}{\partial x_{i}^{2}} &  =-\frac{2}{\delta_{i}^{2}}%
\iint\limits_{\Omega_{i}}\prod\limits_{j\in\mathcal{N}_{i}}\left(
1-p_{j}\right)  dxdy\\
&  +\int_{y_{i}-\delta_{i}}^{y_{i}+\delta_{i}}\frac{2\sqrt{\delta_{i}%
^{2}-\left(  y-y_{i}\right)  ^{2}}}{\delta_{i}^{2}}\prod\nolimits_{j\in
\mathcal{N}_{i}}\left[  1-\hat{p}_{j}\right]  dy\\
&  +\int_{y_{i}-\delta_{i}}^{y_{i}+\delta_{i}}\frac{2\sqrt{\delta_{i}%
^{2}-\left(  y-y_{i}\right)  ^{2}}}{\delta_{i}^{2}}\prod\nolimits_{j\in
\mathcal{N}_{i}}\left[  1-\check{p}_{j}\right]  dy,
\end{align*}
where%
\begin{align*}
\hat{p}_{j} &  =p_{j}\left(  x_{i}+\sqrt{\delta_{i}^{2}-\left(  y-y_{i}%
\right)  ^{2}},y,x_{i},y_{i}\right)  ,\\
\check{p}_{j} &  =p_{j}\left(  x_{i}-\sqrt{\delta_{i}^{2}-\left(
y-y_{i}\right)  ^{2}},y,x_{i},y_{i}\right) ,
\end{align*}%
and the definition of $p_{j}$ is given in~(\ref{sq1}).

Similarly, we have%
\begin{align*}
&\frac{\partial^{2}H}{\partial x_{i}\partial y_{i}} \\  =&\frac{\partial
}{\partial y_{i}}\int_{x_{i}-\delta_{i}}^{x_{i}+\delta_{i}}\int_{y_{i}%
-\sqrt{\delta_{i}^{2}-\left(  x-x_{i}\right)  ^{2}}}^{y_{i}+\sqrt{\delta
_{i}^{2}-\left(  x-x_{i}\right)  ^{2}}}\frac{2\left(  x-x_{i}\right)  }%
{\delta_{i}^{2}}\prod\limits_{j\in\mathcal{N}_{i}}\left(  1-p_{j}\right)
dydx\\
=& \int_{x_{i}-\delta_{i}}^{x_{i}+\delta_{i}}\frac{2\left(  y_{i}-x_{i}%
+\sqrt{\delta_{i}^{2}-\left(  x-x_{i}\right)  ^{2}}\right)  }{\delta_{i}^{2}%
}\prod\nolimits_{j\in\mathcal{N}_{i}}\left(  1-\tilde{p}_{j}\right)  dx\\
& -\int_{x_{i}-\delta_{i}}^{x_{i}+\delta_{i}}\frac{2\left(  y_{i}-x_{i}%
-\sqrt{\delta_{i}^{2}-\left(  x-x_{i}\right)  ^{2}}\right)  }{\delta_{i}^{2}%
}\prod\nolimits_{j\in\mathcal{N}_{i}}\left(  1-\breve{p}_{j}\right)  dx
\end{align*}
where%
\begin{align*}
\tilde{p}_{j}  & =p_{j}\left(  x,y_{i}+\sqrt{\delta_{i}^{2}-\left(
x-x_{i}\right)  ^{2}},x_{i},y_{i}\right),  \\
\breve{p}_{j}  & =p_{j}\left(  x,y_{i}-\sqrt{\delta_{i}^{2}-\left(
x-x_{i}\right)  ^{2}},x_{i},y_{i}\right).
\end{align*}

%


\addtolength{\textheight}{-12cm}


\end{document}